# Trend and Emerging Types of "419" SCAMS

**Falade, Polra Victor**
Cyber Security Department
Nigerian Defence Academy
Kaduna, Nigeria
**E-mail:** poradang@gmail.com, pvfalade@nad.edu.ng
**E-mail:** +2348137205130

## ABSTRACT

Technological advancements have revolutionized various aspects of human life, facilitating communication, business operations, healthcare, education, and environmental monitoring. However, this increased reliance on technology has also led to a surge in cybercrime, including cyber scams. The "419 scam" or Nigerian scam has been a persistent problem for decades, encompassing frauds like advance fee scams, fake lotteries, and black money scams. Initially prevalent through postal mail and later via fax, the scam has now transitioned to email. This study aims to identify recent types of 419 scam emails, particularly after the covid 19 pandemic, and explore commonly used email subjects. Analysis of the sample 419 scam emails revealed trending scams like lucky winner, threat of exposure, business/partnership proposals, investment, cancer/long-term illness, fund, and compensation scams. Emerging scams included COVID-related, cryptocurrency, marketing contact, and software development scams. Irrespective of the scam type, scammers commonly employed email subjects such as 'Re', 'Good day', 'Greetings', 'Dear friend', 'Confirm', 'Attention', and 'Hello dear'. The severity of cybercrime, especially the 419 scams, cannot be overstated, as it erodes trust, causes financial losses, and hampers Nigeria's reputation and economic progress. Combatting cyber scams and enhancing cybersecurity measures are crucial to protect individuals and organizations from falling victim to these fraudulent schemes.

## 1.. INTRODUCTION

The rapid advancement of technology has revolutionized various aspects of human life, enabling activities such as communication, business operations, healthcare management, education, and environmental monitoring (AGBO et al., 2022). Nigeria has experienced a significant increase in internet users, surpassing 100 million in 2019, driven by the widespread adoption of smart devices and improved internet connectivity (NCC, 2019). However, this increased reliance on computers and the internet has also led to a rise in cybercrime, including ATM fraud, phishing, and identity theft, with approximately 14% of Nigeria's internet user population reported to be involved in cybercrime (Ibrahim, 2019).

Cybercrime represents a modern manifestation of traditional criminal acts, leveraging information and communication technology tools and devices (Conteh & Schmick, 2016; Ojilere & Oraegbunam, 2021). It encompasses a wide range of illegal activities facilitated by electronic devices, such as financial fraud, child trafficking, intellectual property theft, and privacy violations through identity theft (Mohammed et al., 2019). Perpetrators of cybercrime exploit computers and internet connectivity to engage in activities like illegal downloading, piracy, and spam campaigns (Hassan et al., 2012).

However, cyber scams are a form of cybercrime that exploit human vulnerabilities using various telecommunications and internet-enabled devices. They target psychological factors like instant gratification, willingness to help, and emotional connection. Victims often struggle to resist persuasion and may become repeat victims. Studies estimate that 10% to 20% of individuals fall prey to scams, with some becoming serial victims. Addressing these vulnerabilities and raising awareness is crucial to mitigate the impact of cyber scams. The progression and sophistication of information technology play a significant role in facilitating these scams and pose risks globally. Nigeria, in particular, faces substantial challenges in dealing with cyber scams (Ojilere & Oraegbunam, 2021; Yoro et al., 2023).

The Nigerian scam, also known as the "419 scam," has been a known problem for several decades. It encompasses various forms of fraud, including advance fee fraud, fake lotteries, and black money scams. The scam initially started with postal mail, then evolved to fax, and eventually transitioned to email. In a 419 scam, the fraudster tricks the victim into paying a certain amount of money with the promise of a larger payoff in the future (Isacenkova et al., 2014). This paper aims at identifying the recent types of 419 scam emails. To determine the recent emerging 419 scams after the pandemic and also identify commonly used 419 scams email subjects.

The paper made the following contributions:
  i.   Collect and evaluate recent samples of 419 scams emails
  ii.  Classify the 419 scam emails into the different types of scams
  iii. Identify the trending and emerging 419 scams
  iv.  Identify the commonly used email subjects in the sample 419 scam emails
  v.   Proffer solutions on how to avoid being scammed

## 2. RELATED WORK

Considerable research has been conducted in the field of 419 scams. The study conducted by Neuhaus (2020) emphasizes the need to understand the tactics employed in famous scam mail narratives and their presentation. This analysis sheds light on the exploitation of human perception, recognition, and thinking, which has become prevalent in the 21st century, where mass media and persuasion dominate various spheres, including politics, public debates, and personal interactions.

Isacenkova et al. (2014) delve into the organization and evolution of scam campaigns using a public dataset. They particularly highlight the role of phone numbers as crucial identifiers for grouping scam messages and provide insights into how scammers operate their campaigns. The study reveals that both email addresses and phone numbers need to appear authentic, often remaining unchanged and reused over extended periods.

Ajayi (2019) adopts ethnographic techniques and draws from Halliday's concept of anti-language to investigate the linguistic strategies employed by cyber scammers in Southwestern Nigeria. Through this approach, the study uncovers linguistic phenomena such as slangy coinages, overlexicalization, relexicalization (semantic extension), and the deployment of sub-strategies such as reduplication, clipping, blending, and acronym usage (initialism). These features characterize the language employed by cyber scammers in the region.

Ojilere & Oraegbunam (2021) contribute to the understanding of cyber scams within the broader context of cybercrimes. Their socio-legal and doctrinal research explores different forms of cyber scams, sheds light on common modus operandi employed by cyber scammers in Nigeria, and highlights the reasons behind individuals engaging in such fraudulent activities. Additionally, the study uncovers the vulnerabilities that cyber scam victims face. The research serves as a critical resource for safeguarding potential victims of cyber scams not only in Nigeria but also in other contexts.

Park et al. (2014) narrowed their focus to Nigerian scams targeting users on Craigslist, a popular online marketplace. Through an in-depth measurement analysis, they aim to comprehend the underground economy of Nigerian scams and identify effective intervention points. By utilizing magnetic honeypot advertisements, designed to attract scammers while repelling legitimate users, the researchers gather scam emails, interact with scammers, and analyze three months' worth of data. Their analysis reveals insights into scammers' action patterns, automation tools, email account usage, and geolocation distribution. Moreover, potential methods to deter these targeted scams are identified based on patterns in scammers' messages and their utilization of email accounts.

Saini (2012) undertakes a specific investigation to identify and analyze the frequently occurring proper nouns within the English text corpus of 419 Nigerian Scams. By employing tokenization and bag-of-words techniques, the study explores country/city names, person names, months, and days of the week. The findings reveal that specific terms like "Africa," "Savimbi," "Friday," and "June" are commonly utilized by scammers and spammers. This analysis contributes to the understanding of the language patterns employed in scam emails.

Collectively, these critical analyses contribute to our understanding of the psychological manipulation employed by scam emails, the organization and characteristics of scam campaigns, linguistic strategies utilized by cyber scammers, the broader context of cyber scams within the domain of cybercrimes, and specific patterns in the text corpus of Nigerian scams. Such insights are essential in combating and protecting potential victims from falling prey to these fraudulent activities. However, there is a need to know the current and emerging 419 scams.

## 3. METHODOLOGY

In this section, we provide a comprehensive overview of the dataset utilized for our analysis of 419 scam emails, along with relevant statistics about the scam messages. The dataset was sourced from the reputable aggregator website https://419scam.org/. This platform serves as a centralized repository for reported scams, gathering data from diverse sources including user reports, dedicated communities, forums, and online activity groups. The selected dataset from 419scam.org encompasses preprocessed data such as email bodies, headers, and specific extracted attributes, including scam categorization and phone numbers. Notably, the dataset does not include IP address information (Isacenkova et al., 2014). It is worth mentioning that 419scam.org organizes the sampled email scams based on the month of the year they were reported or uploaded.

This research paper aims to identify recent and emerging types of 419 scam emails that have been reported. To obtain the most recent reported 419 scam emails, we collected a sample of emails specifically for the year 2023. However, it was challenging to determine the type of 419 scams solely based on extracted features like the email subject. In many cases, the subject line does not provide significant insights into the email's content, and there are numerous instances of repeated subjects. For example, the subject "Attention" was used in multiple emails with varying content. Moreover, scammers employ different subjects and writing structures for the same type of scam. It is possible to observe two emails, potentially from different scammers, constructed differently but with the same intent and a similar story. For instance, the story of a helpless widow confined to a hospital bed, whose deceased husband left behind a substantial fortune intended for charitable purposes, is a common narrative employed by many scammers. However, different types of illnesses, such as various forms of cancer or other long-term ailments, are utilized in these scams.

Given the vast number of reported scam emails available on 419scam.org, manually reading through thousands of emails without any automated tools proved to be a daunting task. Consequently, we opted to limit our analysis to May 2023, the most recent month. We meticulously examined all the scam emails and classified them into different types based on the content and narrative provided by the scammers. The total number of reported email scams for May 2023 amounted to 1381 emails. Out of these, 81 emails were in a language other than English or had no content, containing nonsensical characters, and were therefore excluded from the analysis. Thus, a total of 1300 emails were used for further analysis and classification.

To identify emerging types of scams, we also collected scam emails from May 2019, which was before the pandemic. The 419scam.org archive contained approximately 4,789 scam emails for May 2019. By comparing the scam emails archived before and after the covid 19 pandemic, we determined the emerging scams in the current context.

## 4. FINDINGS

The collected samples of 419 scam emails in this study only capture the initial phase of the scam, known as the baiting phase (Neuhaus, 2020). They do not provide information regarding the success or failure of the scams, as they do not include details on whether the recipients of the emails responded or took any further actions. The purpose of this analysis was to examine the characteristics and types of 419 scam emails based on the provided samples, rather than tracking the complete outcomes or responses from the recipients.

### 4.1 Classification of the Collected 419 Scam Emails
After analyzing the collected reported 419 scams, a thorough examination was conducted by reading and reviewing each archived email. As a result, the scams were classified into 27 distinct types, which include the following: Lucky winner scams, threat of exposure scams, Business/Partnership proposal scams, Supply scams, Dead or alive scams, Compensation scams, COVID-related scams, Next-of-kin scams, order scams, malware/ransomware scams, investment scams, cancer/long-term illness scams, money-in-box scams, loan scams, killed-father scams, fund scams, cryptocurrency scams, fake help scams, authorization scams, foreign affairs scams, contacts for marketing scams, spiritual scams, delivery scams, job offer scams, details scams, romance scams, and software development scams.

### 4.2 Statistics of the 419 Scam Emails
Figure 4.1 presents a graphical representation of the number of 419 scam emails for each scam type. Among the various scam types, four types including lucky winner scams, threat of exposure scams, business/partnership proposal scams, investment scams, and fund scams were reported over 100 times. On the other hand, 12 scams such as authorization scams, foreign affair scams, contacts for marketing scams, spiritual scams, delivery scams, job offer scams, details scams, software development scams, cryptocurrency scams, killed father scams, romance scams, and money-in-box scams were reported less than 20 times. Supply scams, dead or alive scams, compensation scams, COVID-related scams, next-of-kin scams, order scams, malware/ransomware scams, cancer/long-term illness scams, loan scams, and fake help scam were reported between 20 and 100 times. The most frequently reported scam was the business/partnership proposal scam with 195 reports, while the least reported scams were the software development scams, romance scams, and details scams.

Scams types reported from 75 times and above are the trending scams and are briefly described as follows:
- a) **Lucky winner scam:** winner scams operate by deceiving individuals into believing they have won a prize or lottery, or among just luckily selected for a donation even though they never entered such competitions. Scammers typically contact their victims, notifying them of their supposed winnings and requesting payment of a fee or charges to claim the prize. This fee may be disguised as taxes, processing fees, or administrative costs. In reality, there is no prize or lottery win, and the scammers aim to exploit the victims' desire for financial gain.
b) **Threat of exposure scam:** in this type of scam, the scammer asserts that they have hacked into the recipient's system and obtained compromising information about their activities. Hacking involves unauthorized intrusion into a computer, mobile device, or network by individuals seeking to exploit personal data. The scammer claims to possess an incriminating video recording of the recipient's actions and blackmails them by threatening to distribute it to their contacts unless a specified amount of money is paid within a given timeframe.
c) **Business/partnership proposal scam:** in the realm of fraudulent practices, the enduring Nigerian scam manifests as an electronic or written communication, wherein an unscrupulous individual solicits assistance in transferring a substantial sum of money to an international destination. In exchange for providing access to one's personal bank account, the participant is enticed with the prospect of receiving a proportionate share of the funds. Subsequently, the victim is compelled to remit various taxes and fees as a prerequisite for accessing the promised "reward." However, it is imperative to recognize that the promised financial windfall will never materialize, thereby resulting in the irrevocable loss of the fees paid. This deceptive scheme thrives on deceitful inducements and financial manipulation, underscoring the significance of exercising circumspection and discernment when confronted with such fraudulent ploys. This type of scam has the highest number of 419 scam emails.
d) **Compensation scam:** the scammer informs the recipient that they have been chosen for the compensation payment, either as a victim of a scam or due to a delayed payment. They may falsely claim that a repentant scammer is refunding the money or that the government has recovered funds from a scam artist and intends to provide refunds. In some cases, the recipient is asked to pay a fee or provide personal details if they are interested in receiving the compensation.
e) **Investment scam:** investment scams are deceptive schemes crafted by scammers to defraud individuals by offering unrealistic promises of high returns on investments or dubious business ventures. These scams often employ persuasive tactics and false information to entice victims into parting with their money. The scammers may claim to have exclusive knowledge or secret strategies that guarantee significant profits, creating a sense of urgency and excitement. However, the promised returns are typically too good to be true and serve as a way to deceive unsuspecting individuals. The scammer is an investor looking for a good business opportunity to invest in or a person looking to invest his inheritance in a foreign country.
f) **Cancer/ long-term illness scam:** the scammer assumes the role of a childless widow, typically claiming to be the spouse of a wealthy individual from the Middle East or a former diplomat. The narrative revolves around inheriting a substantial fortune from the deceased husband and being afflicted with a life-threatening illness, often breast cancer or a combination of cancer and fibroid problems. The scammer may also mention experiencing a stroke, aiming to evoke sympathy and urgency in the recipient. Nonetheless, the fundamental nature of the scam remains unchanged: the victim is promised access to millions of dollars for minimal effort on their part. However, as the scheme progresses, the victim is inevitably asked to transfer thousands of dollars to an unknown individual posing as a lawyer, security company, or intermediary responsible for managing the funds.

g) **Fund scam:** Fund scams encompass various fraudulent schemes that involve the promise of unexpected money. These scams aim to deceive individuals by offering false inheritances, benefits, or contract payments to obtain their money or sensitive financial information, such as bank or credit card details. The scammer assures the victim that the funds will be transferred through methods such as fund transfers, direct credits, or telegraphic transfers. Additionally, they may claim that the funds have been loaded onto an ATM card, which can be either picked up or shipped to the owner. Subsequently, the scammer requests a fee to facilitate the delivery of the funds.

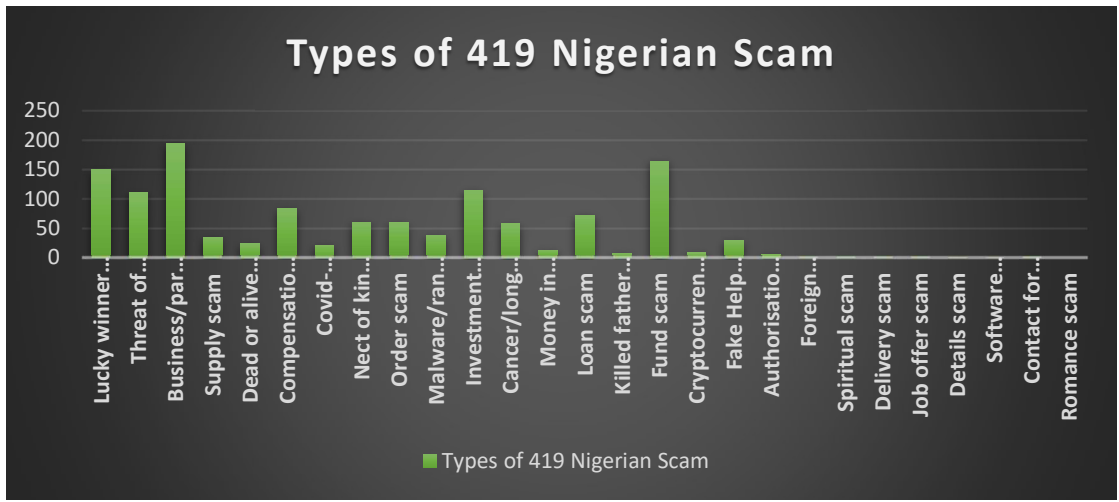

Figure 4.1: Number of reported 419 scamss

### 4.3     Emerging Types of 419 Scams

Based on the analysis of recent case studies, it was identified that there are emerging cyber scams, particularly related to the COVID-19 pandemic. These scams were determined by comparing sample scam emails in May 2019, before the pandemic, with the sample emails of May 2023. The following 419 scams were found to be emerging types: covid-related scam, supply scam, cryptocurrency scam, contacts for marketing scam, software development scam, and threat of exposure scam. These scams were not reported in May 2019, indicating their emergence in recent times.

Among these, the threat of exposure scams had the highest number of reported cases. The supply scam had 35 reported cases, followed by the covid-related scam with 22 reported cases. The remaining scams had less than 10 reported cases each. The covid-related scam is a direct result of the pandemic, while the increase in the use of cryptocurrency has led to the emergence of the cryptocurrency scam. Furthermore, the growing digitalization of businesses and the reliance on technology has contributed to the contact for marketing scams and software development scams.

### 4.4 Commonly Used Subjects of 419 Scam Emails

According to the analysis of recent 419 scam emails, scammers commonly use specific email subjects to manipulate recipients and elicit the desired response. Regardless of the type of scam, the most frequently used email subjects include: 'Re', 'Good day', 'Greetings', 'Dear friend', 'Confirm', 'Confirm receipt', 'Attention', 'Hello dear', 'Acknowledge', 'hello', 'Hi dear', 'urgent', 'morning', 'Good', 'Reply me back', 'reply soon', 'urgent response', and 'Greetings'. Scammers strategically employ these subjects to engage with recipients and prompt them to take actions that benefit the scammers.

For instance, subjects like 'Dear friend', 'Hi dear', and 'hello dear' aim to establish an emotional connection with the recipient. Subjects such as 'good day', 'greetings', 'morning', 'hello', and 'good' are used to display courtesy and create a positive initial impression. In cases where immediate action is required from the recipient, subjects like 'confirm', 'confirm receipt', 'reply', 'reply soon', 'attention', 'acknowledge', 'urgent response', and 'urgent' is employed to create a sense of urgency and prompt a quick response.

### 4.5 Targets of 419 Scam Emails

These 419 scams target victims to obtain two main things from them. These include:
- Victim's data: 419 scam emails aim to deceive victims into providing personal details such as names, addresses, bank account information, phone numbers, and email addresses. Scammers can then use this information for identity theft, hacking, financial fraud, or sending spam messages. Victims should be cautious and avoid sharing personal details with unknown sources to prevent falling victim to these scams and potential harm.
- Victim's money: In most 419 scams, the ultimate goal is to extract money from the victims. Scammers may ask for payment upfront or request fees to be paid in various stages of the scam. The focus is to convince the victims to release their money, either by promising a large sum in return or by creating a sense of urgency or necessity. Individuals must be aware of these tactics and exercise caution to avoid falling victim to financial loss.

### 4.6    How to avoid being scammed
- Caution should be exercised in sending funds to individuals unfamiliar
- Refraining from accessing unsolicited email attachments.
- It is advisable to refrain from disclosing personal or financial information to verify details or receive monetary transactions from individuals with whom you lack familiarity
- It is prudent to exercise caution when encountering unsolicited emails that request monetary transactions or handling
- Any email that proposes the transfer of funds to a third party other than yourself should be met with scepticism and approached with caution.
- It is crucial to verify the legitimacy of investment opportunities and carefully assess the risks involved before committing any funds.

## 5. CONCLUSION

In conclusion, the severity of cybercrime in Nigeria and the world at large, particularly the infamous "419 scam," cannot be overstated. This pervasive criminal activity undermines trust, inflicts substantial financial losses, and impedes Nigeria's reputation and economic development.

Addressing this critical issue necessitates comprehensive and decisive actions, including legislative reforms, enhanced law enforcement capabilities, international cooperation, and intensified public awareness campaigns. Only through these concerted efforts can Nigeria and other nations effectively combat cybercrime, protect potential victims, and restore its standing in the global community.